# Substructure: Clues to the Formation of Clusters of Galaxies

Michael J. West[1,2], Christine Jones[3] and William Forman[3]

## ABSTRACT

We have examined the spatial distribution of substructure in clusters of galaxies using *Einstein* X-ray observations. Subclusters are found to have a markedly anisotropic distribution which reflects the surrounding matter distribution on supercluster scales. Our results suggest a picture in which cluster formation proceeds by mergers of subclusters along large-scale filaments. The implications of such an anisotropic formation process for the shapes, orientations and kinematics of clusters are discussed briefly.

*Subject headings:* cosmology: observations — galaxies: clusters: general— large-scale structure of the universe — X-rays: galaxies

---

[1] Present address: Department of Astronomy & Physics, Saint Mary's University, Halifax, NS B3H 3C3, Canada
[2] Sterrewacht Leiden, Postbus 9513, 2300 RA Leiden, The Netherlands
[3] Harvard-Smithsonian Center for Astrophysics, 60 Garden Street, Cambridge, MA 02138, USA

## 1. Introduction

Substructure appears to be a common feature of many, perhaps most, clusters of galaxies. Current estimates suggest that at least 30% – 50% of rich clusters exhibit multiple concentrations in their galaxy or gas distribution (e.g. Geller & Beers 1982; Dressler & Shectman 1988; Jones & Forman 1992; Mohr, Fabricant & Geller 1993; Salvador-Solé, Sanromà & González-Casado 1993; Bird 1994; Escalera et al. 1994; Stern et al. 1995; see West 1994a for a recent review).

Because dynamical evolution is expected to rapidly erase substructure, its prevalence in rich clusters today strongly suggests that we are currently in the epoch of cluster formation. Quantitative analysis of subclustering may therefore provide information on cluster and galaxy formation, evolution of the intracluster medium and cosmology. For example, a number of authors have suggested that the frequency of substructure in clusters today places a strong lower limit on the cosmological density parameter $\Omega_0$ (Gunn & Gott 1972; Richstone, Loeb & Turner 1992; Kauffmann & White 1993; Lacey & Cole 1993). Similarly, the abundance and mass function of subclusters may yield information about the primordial spectrum of density fluctuations (West, Oemler & Dekel 1988) as well as providing insights to galaxy formation (e.g., Beers & Geller 1983; Dressler 1984; Merritt 1985; Tremaine 1990).

In this paper we endeavor to extract clues about the cluster formation process from the spatial distribution of subclusters. Motivation for this study came from a number of intriguing coincidences that we had noticed between the distribution of subclusters and the surrounding matter distribution on much larger scales. An example is provided by the Coma cluster and its environs shown in Figure 1. Long considered to be the archetype of a rich, relaxed cluster of galaxies, Coma is now known to possess a number of distinct subclusters. The supergiant elliptical galaxies NGC 4889 and NGC 4874 reside in two subclusters in the core of Coma (Fitchett & Webster 1987; Mellier et al. 1988; Davis & Mushotzky 1993; Vikhlinin et al. 1994; Stern et al 1995) and $ROSAT$ observations have revealed a third large subcluster associated with NGC 4839, as well as several other smaller subclusters (Briel, Henry & Böhringer 1992; White et al. 1993). As Figure 1 reveals, the distribution of subclusters within the Coma cluster shows a rather striking alignment with the surrounding large-scale galaxy distribution, in particular the filamentary feature which defines the Coma-Abell 1367 supercluster.

A similar example is provided by Abell 426, which resides in the well-known Perseus-Pisces supercluster. $ROSAT$ observations (Schwarz et al. 1992) show two X-ray peaks which lie right along the prominent supercluster ridge. Another member of the Perseus-Pisces supercluster, the poor cluster AWM 7 (Albert, White & Morgan 1977), also has substructure which shares the same orientation as the supercluster filament (Stern et al. 1995).

These and other examples suggest the intriguing possibility that the distribution of subclusters may be correlated with the surrounding matter distribution on much larger scales. If true, this would be an important clue about the way in which galaxy clusters formed. However a few anecdotal cases like Coma or Perseus are not sufficient to establish the reality of this effect. Therefore, to examine this question more fully we undertook a statistical study of the distribution of substructures in a large sample of clusters.

## 2. The Cluster Sample

X-ray observations have provided a wealth of information on the structure of clusters of galaxies (see Forman & Jones 1990 and Jones & Forman 1992 for reviews). The largest substructure study to date is that of Jones and Forman (1992,1995) who assembled a sample of 366 clusters of galaxies with redshifts less than 0.2 observed with the Einstein satellite. Of these, 208 have adequate signal-to-noise to allow a reliable classification of their X-ray morphologies. Thirty-seven of these show clearly distinguished multiple components, while an additional 56 are elliptical. The subclusters generally have projected separations corresponding to less than $1h^{-1}$ Mpc. It should be noted that other clusters in this sample, particularly the more distant ones, may also have substructure which was not detected owing to the arcminute spatial resolution of the Einstein Imaging Proportional Counter or to the superposition of substructures along the line of sight. For the 93 elliptical or multiple component clusters, we used either the positions of their subclusters or their ellipticity to define a position angle on the sky.

## 3. Linear Subcluster Configurations

Visual inspection of the *Einstein* X-ray images gives a strong impression that when multiple subclusters are present in a cluster they frequently have a collinear distribution. An example is shown in Figure 2. In order to quantify this impression, we performed a simple statistical test using the seven clusters in our sample which have three distinct subcluster components (none of the clusters had more than three subclusters). Three subclusters define a triangle, whose shape can range from isosceles to a straight line. One can define a measure of the triangle shape as

$$S = (D_{max} - D_{int})/D_{min} \qquad (1)$$

where $D_{min}, D_{max}$ and $D_{int}$ are the minimum, maximum and intermediate angular separations between each of the three subcluster pairs. This ranges from $S = 0$ for an isosceles triangle to $S = 1$ for a linear configuration.

We computed this statistic for each of the three-component clusters in our sample; the results are presented in Table 1. Despite the small sample size, the observed distribution of $S$ values differs very significantly from that expected for random subcluster configurations. Monte Carlo simulations were performed of 10,000 random subcluster triplets, and these indicate that the distribution of $S$ values in Table 1 has a probability less than 1% of being consistent with randomly arranged subclusters. Hence there is a strong tendency for linear arrangements of subclusters in clusters. Presumably this is related to the linear shapes of many clusters of galaxies (Rood & Sastry 1971; Struble & Rood 1987).

## 4. Subcluster Orientations with Respect to Large-Scale Structures

One can test the idea that subclusters infall along preferred directions by comparing the orientation of the projected separation vector between each subcluster pair with the surrounding matter distribution on larger scales. Because most clusters are too distant for the surrounding galaxy distribution to be seen in existing surveys, we used Abell clusters to map the surrounding large-scale structure. Although sparser tracers of the large-scale matter distribution than galaxies, rich clusters are known to delineate the same superclusters (e.g., Oort 1983; Bahcall 1992).

For each of the 93 clusters in our sample, the projected position angle defined by its component subclusters or ellipticity, $\phi_{ss}$, was compared with the projected position angle $\phi_{cc}$ from the cluster to each neighboring Abell cluster within a distance $d \leq 10\,h^{-1}$ Mpc (position angles were computed using standard spherical trigonometric relations). The difference between these two position angles defines an acute angle $\theta$,

$$\theta = |\phi_{ss} - \phi_{cc}| \qquad (2)$$

which is a measure of the tendency for the subcluster and cluster pairs to be aligned with one another. Note that there is no ambiguity in assigning a value of $\phi_{ss}$ for clusters with three subclusters, owing to their strongly linear arrangements discussed in the previous section. Spatial separations between clusters were computed assuming a pure Hubble flow with $H_0 = 100\,h$ km s$^{-1}$ and $q_0 = 0.5$. Cluster redshifts were taken primarily from the compilation by Peacock & West (1992) and augmented with a number of other recent measurements (Fetisova et al. 1993; Dalton et al. 1994; Lauer & Postman 1994; Quintana & Ramírez 1995). Figure 3 illustrates the method more clearly, along with the results.

If the orientation of subcluster pairs is independent of the surrounding distribution of Abell clusters, then $\theta$ should be uniformly distributed between 0° and 90°. What one sees in Figure 3 is that for separations less than $10\,h^{-1}$ Mpc the observed distribution is strongly skewed toward small values of $\theta$, which indicates that subcluster pairs tend to share the same orientation as the surrounding large-scale cluster distribution. Of the 93 clusters in our sample, 43 have one or more neighboring clusters within $d = 10\,h^{-1}$ Mpc. A Kolmogorov-Smirnov test confirms the statistical significance of these results; the probability that the distribution seen in Figure 3 could be consistent with a uniform distribution of $\theta$ expected for random orientations of subcluster pairs is only $\sim 0.6\%$. It is certainly significant that the distribution of subclusters appears to "know" about the surrounding distribution of Abell clusters on large scales.

## 5. Discussion

Our results provide clear evidence of a connection between the distribution of subclusters in galaxy clusters and the distribution of neighboring clusters on scales of $\sim 10\,h^{-1}$ Mpc or more.

The most natural interpretation of these results is that cluster formation proceeds by the merging of subclusters along large-scale filamentary features in the matter distribution. This finding is not entirely unexpected, as N-body simulations have shown that filamentary infall may play an important role in some cluster formation models (e.g., West, Villumsen & Dekel 1991; Katz & White 1993; van Haarlem & van de Weygaert 1993; West 1994b, and references therein).

Such an anisotropic formation process has important implications for the shapes, orientations and kinematics of clusters. In particular, this may explain the observed tendency for the major axes of Abell clusters to be aligned with their large-scale environs (e.g., Binggeli 1982; West 1989; Rhee, van Haarlem & Katgert 1992; Plionis 1994; West 1994b). Built up by a series of subcluster mergers which occur along preferred directions, clusters of galaxies will naturally develop orientations that reflect the surrounding filamentary pattern of superclustering. This formation scenario would also lead to strongly anisotropic cluster velocity dispersions.

In conclusion, we have established a connection between the internal structure of clusters of galaxies on sub-Mpc scales and the surrounding large-scale matter distribution on supercluster scales. These results provide an important new clue about the genesis of galaxy clusters, suggesting that cluster formation proceeds via the anisotropic merging of subclusters along filaments.

M.J.W. was supported by the NSERC of Canada and the NFRA of the Netherlands. C.J. and W.F. were supported by the Smithsonian Institution and the *AXAF* Science Center NASA Contract NAS8-39073. We thank the referee, Tina Bird, for her prompt and careful reading of this paper.

TABLE 1
SUBCLUSTER CONFIGURATIONS

| Cluster | S |
|---|---|
| Abell 98 | 0.998 |
| Abell 119 | 0.773 |
| Abell 514 | 0.404 |
| Abell 1477 | 0.999 |
| Abell 1750 | 0.998 |
| Abell 2384 | 0.996 |
| Abell 4067 | 0.988 |

**Figure 1.** Top panel: the large-scale galaxy distribution in the region of the Coma cluster. To highlight features in the galaxy distribution, symbol sizes are proportional to local galaxy density. Circles denote Abell clusters with redshifts $z \leq 0.03$. Bottom panels: the distribution of subclusters in the Coma cluster. Note how the subclusters share the same orientation as the large-scale filament which defines the Coma-Abell 1367 supercluster.

**Figure 2.** *Einstein* X-ray image of Abell 1750, showing the linear arrangement of its three component subclusters.

**Figure 3.** Histogram of the distribution of $\theta$ values for cluster separations $d \leq 10\,h^{-1}$ Mpc, along with a schematic illustration of the method used.